\documentclass[dvips]{article}
\usepackage{amssymb}
\usepackage{amstext}
\usepackage{amsmath}
\usepackage{stmaryrd}
\usepackage{pgf}
\usepackage{color}
\usepackage{pstricks}
\usepackage{array}
\usepackage{rotating}
\usepackage{multirow}
\usepackage{hhline}

\newcommand{\Prob}{{\rm Pr}}

\newcommand{\BBB}{{\cal B}}

\newcommand{\SSS}{{\cal S}}

\renewcommand{\Bbb}{\mathbb}

\newcommand{\RRr}{{\Bbb R}}

\newcounter{countroman}
{\begin{list}{{\rm (\roman{countroman})}}{\usecounter{countroman}}}%
{\end{list}}

\newcounter{countalpha}
{\begin{list}{(\alph{countalpha})}{\usecounter{countalpha}}}%
{\end{list}}

\newcounter{countalphabf}
{\protect\begin{list}{{\rm (}{\bf \protect\alph{countalphabf}}{\rm%
)}}{\protect\usecounter{countalphabf}}}%
{\end{list}}

\mathcode`\<="4268 
\mathcode`\>="5269 
\mathchardef\gt="313E 
\mathchardef\lt="313C 

\newcommand{\beq}{\begin{equation}}
\newcommand{\eeq}{\end{equation}}
\newcommand{\ba}[1]{\begin{array}{#1}}
\newcommand{\ea}{\end{array}}
\newcommand{\bea}{\begin{eqnarray}}
\newcommand{\eea}{\end{eqnarray}}
\newcommand{\bear}{\begin{eqnarray*}}
\newcommand{\eear}{\end{eqnarray*}}

\renewcommand{\paragraph}[1]{\smallskip\noindent {\textbf{#1}.}}

\begin{document}

\title{Towards a Science of Trust
}

\author{
Dusko Pavlovic\\ 
University of Hawaii, Honolulu, USA\\
dusko@hawaii.edu
}

\date{}

\maketitle

\begin{flushright}
\parbox{7.2cm}{{\it \footnotesize It is unclear how to think about trust or to model its ebb and flow. Is there some sort of Second Law of Thermodynamics of trust, where trust starts high and is dissipated over time? Or is it the contrary, that trust starts low and can grow through a series of good experiences? Is it more complex, and how can the waxing and waning be thought about?}}\\[3ex] 
{\footnotesize JASON Report
on Science of Cyber-Security \cite{jason2010SoS}} 
\end{flushright}

\begin{abstract}
The diverse views of science of security have opened up several alleys towards applying the methods of science to security. We pursue a different kind of connection between science and security. This paper explores the idea that security is not just a suitable \emph{subject}\/ for science, but that the process of security is also \emph{similar} to the process of science. This similarity arises from the fact that both science and security depend on the methods of \emph{inductive inference}. Because of this dependency, a scientific theory can never be definitely proved, but can only be disproved by new evidence, and improved into a better theory. Because of the same dependency, every security claim and method has a lifetime, and always eventually needs to be improved.

In this general framework of \emph{security-as-science}, we explore the ways to apply the methods of scientific induction in the process of trust. The process of trust building and updating is viewed as hypothesis testing. We propose to formulate the trust hypotheses by the methods of algorithmic learning, and to build more robust trust testing and vetting methodologies on the solid foundations of statistical inference.
\end{abstract}


%
%

\section{Introduction}

The effort towards science of security was born from the need for a more systematic approach to security \cite{jason2010SoS,MeushawR:TNW12,LandwehrC:TNW12,SchneiderF:TNW12,PavlovicD:TNW12}. It resulted in new  empiric and experimental approaches to cyber security \cite{Benzel:ACSAC11,hard-problems,HotSoS14}. The fact that science of security still means many things to many people should perhaps be seen as a feature and not a bug, since already security on its own means many things to many people, and it is natural that they study it from many directions \cite{PavlovicD:TNW12}. On the other hand, it seems that each step of scientific progress requires a unifying idea, each of them showing that a certain group of trees is actually a forest \cite{KuhnT:structure}. What is then the unifying idea of science of security?

\subsection{Science is something else}\label{Sec:Science}
\subsubsection{What science is not} 

Every known civilization seems to have developed technology, art, and religion. But  only the Western Civilization has developed science. Science emerged in Europe during the Renaissance, and caused the Industrial Revolution. This unique stream of events is analyzed in some detail in \cite{KuhnT:structure}.\footnote{It has been objected that this view can be construed as eurocentric. While the word \emph{"science"} can, of course, be used to denote many things, as explained in the next paragraph, theory of science \emph{defines}\/ science as the movement that led to the Industrial Revolution. Since the fact that the Industrial Revolution emerged in Europe is historically uncontestable, the fact that science emerged in Europe follows from this definition. Moreover, as incisive critics of the Industrial Revolution even before its current destructive consequences became clear, the theorists of science can  hardly be accused of \emph{praising}\/ Europe for being the cradle of science \cite{Feyerabend}.} 

There are, of course, many definitions of science. Some of them are shaped to include the teachings of Ron L.~Hubbard; some to include marxism, or even the daily thoughts of the current leader of North Korea. Most definitions, however, point to some of the features of the methodological movement that led to understanding the natural processes like heat, electricity, magnetism, radiation, or networking. Although the notion of science can be extended to include astrology, scientology, theology, mathematics, or engineering, it does not seem useful to stretch it too much. Assigning the status of science, say, to the engineering principles and processes (whether those that enabled the public works of Ancient Egypt, or those that emerged in medieval alchemy, or in Renaissance architecture, or in modern software engineering) might conceal something essential about science. Can science be reduced to its technological thrust \cite{LandwehrC:TNW12}? Or does it boil down to the view that the world is governed by a system of laws \cite{SchneiderF:TNW12}? Or is there more to it? 

Many ancient civilizations developed the \emph{quantitative methods}\/ that enabled them to plan and execute extensive engineering projects, and change the landscapes of their environment. Many of them also explained the world around them through sophisticated \emph{theoretical edifices}\/ and that included the \emph{'Laws of Nature'}, formalized as mythologies, or gathered in sacred texts, often equipped with extensive symbolic systems. But no one until the Age of Science came anywhere near to understanding and reproducing, e.g., the thermo-nuclear processes of Sun; or the space-time curvature, without which our GPS systems could not surf on the geodesics, and would keep sending us wrong coordinates. No one before science came anywhere near to understanding genomics and to engineering the basic processes of life; and nowhere near to connecting our world into a network of networks, and spanning a distance-free space, where every two nodes are neighbors, and where our joint problem solving, and problem creating capabilities seem to be reaching a completely new level. This network of networks is what we call cyber space. Inhabited by the processes that we programmed, but whose interactions we cannot control, cyber space hosts  a new nature in need of a new science. Is this new level of our civilization just another new level of yet another civilization, or is it something else? Is the science that brought it all about just another way that we found to generate new technologies, or just another religion that tells us the laws of the world, or is it something essentially new?

There is a qualitative difference between the science-ge\-ne\-ra\-ted technologies, and the spontaneously evolved technologies. There is also a qualitative difference between the symbolic systems of theologies and mythologies that emerge from religions, and the symbolic systems of mathematics and computation that underlie science. There is a qualitative difference between the religious rituals on one hand and the scientific protocols on the other. The essence of these differences is not in the levels of complexity or effectiveness. There are complex religious systems, and there are simple scientific theories. Many religions and even superstitions postulate their \emph{'Laws of Nature'} that are structurally indistinguishable from those postulated by science. Astrology and phrenology have in their time been tested as scientific theories, by scientific methods, and rejected not for structural reasons, not as unscientific, but as wrong. And there are also effective religious systems, and there are ineffective sciences. E.g., although the processes of photosynthesis are everywhere around us, at the bottom of all of our food chains, science has remained unable to understand what do the plants really do when they bind photons into sugars. There is a quantum effect, but science has been ineffective in explaining it. It has also been less effective than most religions in addressing people's emotional and social needs. 

So what really distinguishes scientific theories, if it is not complexity, and not effectiveness? 

\subsubsection{What science is}

I propose to consider the logical pattern of inductive inference as the essence of science: \emph{While religion claims to provide the truth, science only seeks to disprove false hypotheses.}\footnote{There are, of course, many other ways to characterize science. The claim here is that this one is useful for the purposes of science of security.}

In a formal sense, science is the quest for \emph{disproving}\/ theories.   
This formal sense was fully implemented for the first time in Ronald Fisher's practical methods of scientific inference \cite{FisherR:1925,FisherR:1959}, and then analyzed theoretically in Karl Popper's extensive and influential work \cite{PopperK:logic}. The historic support for this view of science was provided by Thomas Kuhn \cite{KuhnT:structure}, while the scientists themselves provided some of the most compelling examples from their current practices \cite{FeynmanR:character}.  Other leading templates of scientific inference (e.g. the Neyman-Pearson testing \cite{Neyman:Pearson:1933}, or Bayesian inference \cite{Bayes,Bernardo-Smith:bayesian}) may appear to offer ways beyond this negative logic of science, as the quest for merely \emph{improving}\/ scientific theories through \emph{disproving}\/ false hypotheses. But a closer look shows that they only formalize the task of hypothesis selection, and thus support formation of new theories, not proving. They do not provide a method to definitely prove anything. Richard Feynman announced this with compelling simplicity in his lectures on \emph{'The Character of Physical Law'} \cite{FeynmanR:character}:
\begin{quote}
If we have a definite theory, from which we can compute the consequences which can be compared with experiment, then in principle we can prove that theory wrong. But notice that we can never prove it right.  Suppose that you invent a theory, calculate the consequences, and discover every time that the consequences agree with the experiment. The theory is then right? No, it is simply not proved wrong! In the future you could compute a wider range of consequences, there could be a wider range of experiments, and you might then discover that the thing is wrong. [\ldots] --- 
%
\emph{We never are definitely right; we can only be sure when we are wrong.}
\end{quote}
This is perhaps the best kept secret of science: \emph{Science does not provide persistent theories; it only provides methods to disprove and improve our hypotheses.}

\subsection{Security is like science}

The fact that the process of security is of the same type like the process of science can be illustrated by translating Feynman's statement from the language of science to the language of security:
\begin{quote}
If we have a precisely defined security claim about a system, from which we can derive the consequences which can be tested, then in principle we can prove that the system is insecure. But we can never prove that it is secure. Suppose that you design a system, calculate some security claims, and discover every time that the system remains secure under all tests. The system is then secure? No, it is simply not proved insecure! In the future you could refine the security model, there could be a wider range of tests and attacks, and you might then discover that the thing is insecure. --- \emph{We never are definitely secure; 
we can only be sure when we are insecure.}
\end{quote}

A scientific approach to security must therefore begin with the realization that there is no persistent security. Cryptographers have known for a long time that every key has a lifetime. It is time that we recognize that every security claim has a lifetime. The designers of protocols and systems have, of course, accumulated a lot of empiric evidence about this phenomenon \cite{PavlovicD:SEFM10}. The point is to understand it as a \emph{logical}\/ phenomenon.

Upon the admission that theories cannot be definitely proved, but only disproved and improved, science has gained its current unparalleled power to harness nature. Upon the realization that security guarantees cannot be definitely assured, but only broken and strengthened, science of security will gain the ability to tap its power to protect from the same methodological source.

\subsection{Zoom on trust}
In this paper we focus on the scientific approaches to a special family of security claims: the statements of \emph{trust}. While a general security claim says that a key $K$ is uncompromised, or that a protocol $P$ guarantees an authentic channel, a statement of trust says that Alice trusts the key $K$ for use in a particular cipher, or that Bob trusts the protocol $P$ to establish an authentic channel with Alice. A statement of trust is thus a security statement \emph{bound}\/ to two subjects and an object: \emph{who trusts what to whom}.  
The parallel between the security processes of trust building and the scientific methods of hypothesis testing seems like a particularly good illustration of the general logical link of security and science, so we pursue it in the rest of this paper.

\subsection*{Outline of the paper}

In Sec.~\ref{Sec:Trust} we briefly explain the concept of trust used in the paper, and why is it interesting to model the process of trust as hypothesis testing. In Sec.~\ref{Sec:Testing} we show on toy examples how to apply the three standard methods of statistical inference in trust testing. In Sec.~\ref{Sec:Formulating} we show how to formulate the best trust hypotheses \emph{a priori}, since it is notoriously difficult to extract the normal behavioral profiles from empiric data. In Sec.~\ref{Sec:Final} we comment about the relations of the presented ideas with the other views of trust, and with the application of statistics in intrusion detection.

\section{Trust as hypothesis testing}\label{Sec:Trust}

\subsection{What is trust?}\label{sec-what-trust}

Security analyses often begin with the assumptions that some of the subjects are \emph{honest}, i.e. that they behave according to some prescribed protocol rules, whereas the others are dishonest, and launch attacks. Trust internalizes the honesty assumptions into beliefs of subjects about each other. E.g., we say that Alice trusts Bob if she believes that he will behave honestly, according to some protocol agreed implicitly or explicitly. In such a trust statement, Alice is the \emph{trustor}, and Bob is the \emph{trustee}. In social and electronic networks, and on the web, trust is implemented in a variety of ways: as feedback services in web commerce, as the web of trust or certificate authorities in the various versions of Public Key Infrastructure, etc. The underlying trust models often include \emph{trust ratings}, which quantify trust, and the \emph{entrusted concepts}, which qualify trust. A survey of the models of trust used in computer security research can be found in \cite{BoydC:trust-survey}. Dynamics of the trust processes  in network computation were analyzed in \cite{NicolD:IEEE10,PavlovicD:FAST08,PavlovicD:FAST10}, and the problem of trust was introduced in the framework of science of security in \cite{NicolD:HotSoS14}. 

\subsection{Inductive inference of trust}

Just like science can never settle but has to keep testing its theories and refining its hypotheses, trust can also never settle and needs to keep testing its hypotheses. Just like a scientific theory can always turn out to be wrong, trust can always be broken. The reasoning about such ongoing processes goes under the name of inductive logics, which is quite different, and much less familiar than deductive logics. The central problem of the inductive inference of trust is expressed by the central principle of the modern court of law, i.e. the principle of \emph{due process}: that the accused must be \emph{presumed innocent until proven guilty} \cite{PenningtonK:innocent}. But this is just the legal form of a more general social principle of trust: that people should be \emph{trusted until proven untrustworthy}.The  burden of proof is here on the prosecution, or on the accusers.  The dual principle of \emph{ordeal}, typical of medieval trials, places the burden of proof on the defense, and requires that the accused be \emph{presumed guilty until proven innocent}. The corresponding social maxim is the principle of distrust (or suspicion), namely that people should be \emph{trusted only if they are proven trustworthy}. These two views of trust, the optimistic and the pessimistic one, correspond to the two social functions of trust:
\begin{itemize}
\item to support stable social links based on \emph{cumulative trust}: "I trust you because I know you"
\item to enable new social links through a \emph{leap of trust}: "I trust you although I don't know you"
\end{itemize}
Note that both the trust principle and the suspicion principle are asserted in a logical process akin to science: they are hypotheses that need to be tested. The logical parallel described in the Introduction emerges again: just like a scientific theory can always be disproved by a new experiment, but can never be definitely proven, trust can always be broken, and can never be settled. We just follow this parallel.

%

\subsection{How to trust methodically?}

The scientific method is the method for hypothesis testing through empiric validation. This means that a scientific theory can only be validated on a finite number of samples or instances, since the empiric data are always finite. Hence the asymmetry of inductive inference: while a counterexample can definitely disprove a theory, no amount of experience can definitely prove it. This is where the \emph{problem of induction} emerges \cite{induction-lakatos}.

Statistical methods have been developed as tools for deciding when to reject a hypothesis \cite{FisherR:1925,FisherR:1959}, and also which alternative hypothesis to endorse \cite{Neyman:Pearson:1928,Neyman:Pearson:1933}. In the experimental setting, statistical methods moreover allow testing multiple hypotheses and quantifying their likelihood \cite{CoxDR:theoretical}.

\subsection{How many trust values?}
Up to the point where the trust \emph{decisions}\/ need to be made, trust can be quantified in many ways, reflected to some extent by the trust ratings, as mentioned in Sec.~\ref{sec-what-trust}. There may be many colors, shades, and nuances of trust, in-between trusting and not trusting. At the end of the day, though, a trust decision must be extracted: \emph{Will the trustor trust the trustee enough to enter into the entrusted transaction?}\/ At the moment of decision, all previous considerations are reduced to one of the two answers: \emph{yes}\/ or \emph{no}. This simple outcome is not only the \emph{process requirement}\/ of trust, akin to the process requirement of justice, where the verdict of \emph{guilty}\/ or \emph{not guilty}\/ must be extracted from whatever mixture of subtle and dubious concerns may precede it. More importantly, the final trust decision is in principle also the \emph{only}\/ observable manifestation of trust. The rich models of trust are our theories, attempting to explain the unobservable causes of the trust decisions. With such theories, science always does the same thing: it tests them as hypotheses, and decides whether they should be rejected or not yet. The good news is that the trust process seems similar. The bad news is that the \emph{yes-no}\/ decisions are not simple.

In Sec.~\ref{Sec:Testing}, we sketch how the basic statistical methodologies apply to trust decisions, i.e. how the trust hypotheses can be tested scientifically. In the subsequent Sec.~\ref{Sec:Formulating}, we discuss a harder problem of trust science, that does not yield to the standard methodologies: \emph{how to formulate the trust hypotheses for testing}.

\section{Testing trust hypotheses}\label{Sec:Testing}
Suppose that you are interacting with a system $\SSS$ presented by a set of observable behaviors $\BBB$. Depending on the ongoing observations of the system behaviors, you must make decisions whether to entrust the system with some critical or security sensitive operations. For instance, if $\SSS$ is a computational device, then $\BBB$ can consist of the various computational behaviors: it may run fast or slowly, it may crash or spontaneously restart, it may show high or low CPU load, frequent or intermittent network accesses, various power usage behaviors, etc. If $\SSS$ is a closed network or a large organization, then the observable behaviors $\BBB$ may consist of the various network phenomena, such as local load imbalances, clustering and community formations, network chatter or its absence, and so on. If $\SSS$ is a market segment or a network of contractors, then $\BBB$ consists of the various market behaviors: clear or unclear market positions and strategies, pricing drift, shifts in supply or demand, overt or covert information propagation. In all cases, it is interesting to assume that the observable behaviors conceal some ultimately unobservable causes:  the computational device may have a firmware virus or a hidden hardware component; the organization may be penetrated by undetectable moles, or bubbling with defectors; the market may be manipulated by a colluding cluster, or swayed by hidden incentives. --- \emph{Science offers methods to detect the unobservable causes of some observable phenomena.}

The observations of the observable behaviors $\BBB$ are modeled by a real function $f:\BBB\to \RRr$, which is often called a \emph{statistic}. A statistic may list the raw measurements of a sample, but it more often displays some property, e.g. the mean, the deviation, a higher-order moment, or some other combination of data. 

One thing that a statistic does not display is a \emph{distribution}\/ of the behaviors in $\BBB$. The distribution of the behaviors, i.e. how often does a behavior $b\in \BBB$ come about in a system $\SSS$, is what a scientific analysis attempts to induce from the observations. More precisely, a scientific analysis proceeds by
\begin{itemize}
\item[(1)] setting a hypothesis $\theta$, presented by a probability distribution $\Pr_\theta :\BBB \to [0,1]$, and then
\item[(2)] testing whether the statistic $f:\BBB\to \RRr$ supports or disproves the hypothesis $\theta$.
\end{itemize}
In the context of trust, the probability distribution $\Pr_\theta :\BBB \to [0,1]$ is intended to capture the trust profile of the system $\SSS$: e.g., how often does it manifest the undesirable behaviors, how reliable is its track record, etc. Testing the trust hypothesis $\theta$ should tell us whether to stick with it, or replace it with another trust statement.

In this section, we assume that the trust hypothesis $\Pr_\theta$ is given: e.g. from the records of past behaviors. The statistic $f$ presents a new record, capturing recent behaviors. The task is to align the two. The problem of formulating $\Pr_\theta$ will be discussed in the next section.

\subsection{Significance testing of trust}\label{Sec:Significance}
For simplicity, assume that the system $\SSS$ has just 4 observable behaviors, collected in the set $\BBB = \{a,b,c,d\}$. To be trustworthy, the system should manifest the acceptable behavior $a$ at least $98\%$ of time. It may block $b$, or crash $c$ for $.5\%$ of the time, and it may delay $d$ its functioning for $1\%$ of the time. So we postulate the \emph{null hypothesis}\/ that the system $\SSS$ behaves according to the probability distribution  $\Pr_0: \{a,b,c,d\}\to [0,1]$ displayed on Table~\ref{Tab:null}.
\begin{table}[htdp]
\begin{center}
\begin{tabular}{|r||c|c|c|c|}
\hline
$\BBB$ & $a$ & $b$ & $c$ & $d$ \\
\hline
\hline
$\Prob_0$ & .98 & .005 & .005 & .01\\
\hline
\end{tabular}
\caption{Trustworthy behavior}
\end{center}
\label{Tab:null}
\end{table}%
For even more simplicity, assume that we observe just one of the events from the set $\{a,b,c,d\}$. This means that the statistic $f:\{a,b,c,d\}\to \RRr$ will have the value 1 for one event, and 0 for the rest. Should we continue to trust the system $\SSS$?

In statistics, the answer to this question is reduced to determining whether the sample represented by the statistic $f$ is \emph{significant}\/ enough to reject the null hypothesis (which was in our case that the system $\SSS$ was trustworthy). The idea of statistical \emph{significance testing}\/ is that the observation $f$ is significant enough to reject the null hypothesis just when the observation $f$ is extremely unlikely according to the null hypothesis. So we could fix a very small number $\alpha \gt 0$ and say that the null hypothesis should be rejected if $x$ is observed such that
\bea
\Prob_0\big(f(x) = 1\big) &\lt & \alpha
\eea
Since the times before computers, the scientists got in the habit of tabulating and using $\alpha = 5\%$ and $\alpha = 1\%$. So if we use $\alpha = 1\%$ and observe $b$ or $c$, we would have to reject the null hypothesis, and stop trusting the system $\SSS$; and if we observe $a$ or $d$ we could continue to trust it.

But to not oversimplify things, we should mention that already the founder of statistics, Ronald Fisher, argued in \cite{FisherR:1925,FisherR:1959} that a test should be considered significant and the null hypothesis rejected only when
\bea\label{pval}
\sum_{\Prob_0(y) \leq P}\Prob_0(y)
& \lt & \alpha
\eea
where $P = \Prob_0\big(f(x)=1\big)$ for the observed event $x$. In words, the total probability of all events $y$ that are at least as unlikely as the observed event $x$ should be less than $\alpha$. The left-hand side of \eqref{pval} is the \emph{$p$-value} of the observation $f$ under the hypothesis $\Prob_0$. The $p$-value of both $b$ and $c$ is now $.1$, and the null hypothesis is never rejected. The $p$-values for $a$ and $d$ are $1$ and $.2$ respectively. 

\paragraph{Remark} It should be noted here that significance testing is a typical embodiment of the \emph{negative}\/ logics of scientific induction: a test is only significant if it \emph{disproves}\/ the null hypothesis. This aspect of inductive logic is similar to the proof by contradiction in deductive logic; but it is different from deductive logic in that this is the \emph{only}\/ inductive proof schema, while deductive logic also has the positive schemas. This logical constraint is \emph{just}\/ what makes inductive logic and the scientific methodologies built upon it, suitable for the reasoning about security and trust, as it echoes the fact that they can always be broken, and cannot be assured by logics.

\subsection{Powerful testing of trust}\label{Sec:Pow}
While the significance testing allows rejecting the null hypothesis when significant tests are found, it does not allow drawing any conclusions about the null hypothesis when it is not rejected, and no conclusions about the other hypotheses when the null hypothesis is rejected. The testing method devised by Neyman and Pearson \cite{Neyman:Pearson:1928,Neyman:Pearson:1933} considers two competing hypotheses $\Prob_\theta:\BBB\to [0,1]$, for $\theta\in \{0,1\}$, and maximizes the probability that the null hypothesis $\theta = 0$ is rejected when the alternate hypothesis $\theta = 1$ happens to be true. This probability is called the \emph{power}\/ of a test. 

It is assumed that the null hypothesis $\theta = 0$, claiming that the observed sample will be distributed according to $\Prob_0:\BBB \to [0,1]$, is the one that is currently accepted, whereas the alternate hypothesis $\theta = 1$, claiming that the observations will be distributed according to $\Prob_1:\BBB \to [0,1]$, will gain validity if the test turns out to be significant and rejects the null hypothesis. For instance, when a scientist hypothesizes that a phenomenon $A$ is the cause of the phenomenon $B$, then the null hypothesis is usually taken to be the claim that the phenomenon $B$ is not correlated to $A$, whereas the alternate hypothesis is the claim $A$ and $B$ are correlated. When a judge needs to decide whether the accused $A$ has committed a crime $B$, then the null hypothesis is that $A$ is innocent with respect to $B$, whereas the alternate hypothesis is that $A$ is guilty of $B$.

To continue with the example from Sec.~\ref{Sec:Significance}, now consider the two hypothetic distributions of the behaviors in the system $\SSS$ displayed in Table~\ref{Tab:alt}. In the last line of the table is the \emph{likelihood ratio}\/ $\frac{\Prob_1(x)}{\Prob_0(x)}$. Neyman and Pearson \cite{Neyman:Pearson:1928} use the likelihood ratio to decide when to reject the null hypothesis $\theta = 0$ in favor of the alternative hypothesis $\theta = 1$.
\begin{table}[htdp]
\begin{center}
\begin{tabular}{|r||c|c|c|c|}
\hline
$\BBB$ & $a$ & $b$ & $c$ & $d$ \\
\hline
\hline
$\Prob_0$ & .98 & .005 & .005 & .01\\
\hline
$\Prob_1$ & .098 & .001 & .001 & .9\\
\hline\hline
$\frac{\Prob_1(x)}{\Prob_0(x)}$ & .1 & .2 & .2 & 90 \\
\hline
\end{tabular}
\caption{Trustworthy vs untrustworthy behavior}
\end{center}
\label{Tab:alphabeta}
\end{table}%
 For this purpose, they introduce the decision thresholds $\alpha$ and $\beta$, displayed in Table~\ref{Tab:alphabeta}, which define the error probabilities as follows
\begin{itemize}
\item $\alpha$ is the probability that the null hypothesis is rejected when it is true, whereas
\item $\beta$ is the probability that the null hypothesis is not rejected when it is false.
\end{itemize}
\begin{table}[htdp]
\begin{center}
\begin{tabular}{cr||c|c|}
\cline{3-4}
&&  \multicolumn{2}{c|}{reality} \\
\cline{3-4}
 & & $\theta = 1$ & $\theta = 0$ \\
\hline\hline
\multicolumn{1}{|c|}{\multirow{2}{*}{decision}} & $\theta = 1$ & \begin{minipage}[c]{2.5cm}
true \\
$1-\alpha$ confidence
\end{minipage} &\begin{minipage}[c]{2.5cm}
false negative \\
$\beta = \Pr(0|1)$
\end{minipage} \\
\cline{2-4}
\multicolumn{1}{|c|}{} & $\theta = 0$ & \begin{minipage}[c]{2.5cm}
false positive \\
$\alpha = \Pr(1|0)$
\end{minipage} & \begin{minipage}[c]{2.5cm}
true \\
$1-\beta$ strength
\end{minipage}
\\
\hline
\end{tabular}
\caption{Decision thresholds $\alpha$ and $\beta$}
\end{center}
\label{Tab:alt}
\end{table}%
Since the rejection of the null hypothesis is conventionally viewed as the positive outcome a statistical test, the first type of error is called a \emph{false positive}\/ decision, whereas the second type of error is called a \emph{false negative}. E.g. in the court of law,  sentencing an innocent person is a false positive, and letting a guilty person free is a false negative, since the null hypothesis is that the accused is innocent, and the burden of proof towards rejecting this hypothesis is on the prosecution. In a fire alarm system, the null hypothesis is that there is no fire, and the false positive is when the alarm rings without fire, whereas a false negative is when the alarm does not ring when there is fire. It is generally accepted as worse to have false positives, since they lead to switching off the fire alarms, rejecting the entire testing frameworks, and thus impelling the negatives as the only outcomes. Neyman and Pearson therefore design the testing frameworks where the upper bound $\alpha$ of the false positive decisions is chosen by the tester, and then the upper bound for the false negative decisions is minimized. The \emph{power}\/ of a test is defined to be the probability $1-\beta$ that the null hypothesis is rejected when it is really false. The Neyman-Pearson Lemma \cite{Neyman:Pearson:1928} says that the maximally powerful test is given by the decision rule that the null hypothesis of innocence should be rejected whenever the likelihood of guilt is
\bea\label{eq-NP}
L(x) \ =\ \frac{\Pr_1(x)}{\Pr_0(x)} & \gt & \eta
\eea
where the threshold $\eta$ is such that the chance of false positives is
\bea
\Pr\left(L(x) \gt \eta\ |\ \theta = 0\right) & =  &\alpha
\eea
The claim that \eqref{eq-NP} gives the most powerful test means that if the chance $\alpha$ in \eqref{eq-alpha} is fixed, then $\beta$ in
\bea
\Pr\left(L(x) \leq \eta\ |\ \theta = 1\right) & = & \beta
\eea
is minimal for the fixed when $L(x) = \frac{\Pr_1(x)}{\Pr_0(x)}$. Recall that $\alpha$ is the chance that an innocent subject is found guilty, whereas $\beta$ is the chance that a guilty subject is found innocent. Going back to the trust test from Sec.~\ref{Sec:Significance}, where $f:\{a,b,c,d\} \to \RRr$ captured the observation of a single system event, the Neyman-Pearson powerful testing would reject the null hypothesis $\theta = 0$ in favor of the alternative hypothesis $\theta = 1$ at the level $\alpha = 1\%$ only if the event $d$ is observed, and otherwise fails to obtain a significant result. This means that we should only reject the trust hypothesis $\theta = 0$ and endorse the hypothesis $\theta = 1$ that the system $\SSS$ is not trustworthy if the observed delays $d$ amount to more than $1\%$ of the sampled performance time. Crashing or blocking $.5\%$ of the time should not trigger our distrust. 

Note that the threshold $\alpha = 1\%$, imposed in the powerful testing as the upper bound of the false positives, has eliminated the significance of the observations $b$ and $c$, which were significant enough to cause the rejection of the null hypothesis at the same threshold level $\alpha = 1\%$ in Sec.~\ref{Sec:Significance}. On the other hand, the minimization of the false negatives in the powerful testing has now introduced the observation $d$ as significant, which it was not the significance testing. The two testing approaches thus implement two incomparable views of trust. It seems worth while to further explore which one might be more suitable for which application domains.

Although the powerful testing allows comparing pairs of hypotheses (albeit in essentially asymmetric roles of the null hypothesis and its alternative!), it actually provides little help in selecting between \emph{multiple}\/ alternative hypotheses. The best we can do with powerful testing in such situations is to test the null hypothesis against each of the candidate alternatives. However, such approaches lead to pathological situations, where the hypothesis $0$ is rejected against $1$, $1$ against $2$, and $2$ against $1$. Similar phenomena arise when the same significance test is applied to several hypotheses, in the hope that some will be rejected and some not. Overcoming such difficulties requires randomized sampling, Bayesian reasoning, and controlled experiments.

\subsection{Experimental testing of trust}\label{Sec:Exp}
If I know an overall probability $\Pr(0)$ that a system similar to $\SSS$ might be trustworthy, and $\Pr(1) =1-\Pr(0)$ that it might not be trustworthy, then I could derive the probability $\Pr(0|x)$ that the system $\SSS$ is trustworthy after the observed behavior $x\in \BBB$ using the Bayes' law:
\bea
\Pr(0|x) & = & \frac{\Prob_0(x)\Pr(0)}{\Prob_0(x)\Pr(0) + \Prob_1(x)\Pr(1)}
\eea
If there are several hypotheses $\theta \in \Theta = \{0, 1, 2,\ldots, n\}$ about the behavioral profiles of the systems, then I can calculate the probability of each of them after the observation $x\in \BBB$ by the general formula
\bea\label{Bayes}
\Pr(\theta |x) & = & \frac{\Prob_\theta(x)\Pr(\theta)}{\sum_{\psi \in \Theta} \Prob_\psi(x)\Pr(\psi)}
\eea
However, the only way to control the distribution $\Pr: \Theta \to [0,1]$ of the trust profiles of a population of systems to which $\SSS$ belongs is to model this population in the experimental environment of a laboratory, where I could control that the sample is distributed according to $\Pr: \Theta \to [0,1]$. Sampling the behaviors of the system $\SSS$ in this controlled environment would then allow me to calculate $\Pr(\theta | x)$ according to \eqref{Bayes} for all profiles $\theta \in \Theta$, and to select the most likely profile $\theta = 0\in \Theta$ as my current trust hypothesis about $\SSS$.

But even this experimental environment, where I can impose the prior probability $\Pr:\Theta \to [0,1]$ by controlling the sample, does not give me the prior probabilities $\Pr_\theta :\BBB\to [0,1]$, which express the trust hypotheses to be tested. Where do they come from?

\section{Formulating trust hypotheses}\label{Sec:Formulating}

How exactly should I find the trust hypotheses suitable for testing? How should I select the most important ones? 

\subsection{The scientific presumption of innocence}\label{Sec:NullTrust}
Both the scientific methodology and the sound legal practices suggest that the null hypothesis should be that the system is trustworthy, i.e. "innocent until proven guilty" \cite{PenningtonK:innocent}. The alternate hypotheses should describe the various forms of undesired behavior, which the tested sample might uncover if the null hypothesis is rejected. 

If I know the statistical profile of the desired normal behavior of a system, then I should take that profile as the null hypothesis $\Prob_0:\BBB\to [0,1]$. But it is usually difficult to specify the desired normal behavior as a single profile. It is much easier to characterize each of the abnormal behaviors, which we learn from the anomalies experienced in the past. That is why the statistical intrusion detection systems \cite{DenningD:IDM,LuntT:IDS,IDS-Bayes}  and forensics mostly work with the statistical profiles of intruders and criminals, and test these profiles as the null hypotheses. 

The problem with this "guilty until proven innocent" approach is not just that it is unfair in court. A greater problem arises from the logical limitation of inductive inference: that \emph{the null hypothesis can never be proved by a finite number of tests, but can only be disproved}. By testing the profiles of guilt on the given samples of behaviors, we can never demonstrate anyone's guilt; we can only fail to disprove it. The consequence in the realm of security is that the trust based on testing and rejecting every known form of undesirable behavior is not only impractical, but also the weakest possible form of trust. All that you know is that no guilt has been proven yet. The complexity and the ineffectiveness of this method is illustrated time and again by the complexity and the ineffectiveness of the vetting procedures, which often admit untrustworthy subjects, while regularly rejecting trustworthy subjects. Scientifically based trust, based on testing the null hypothesis that the subject is trustworthy, would obviously be  simpler and more effective, both because it allows sound statistical controls of the false positives and the false negatives, and also because it eliminates not only the known anomalies, but all anomalies that are inconsistent with the normal behavior profile described by the null hypothesis.

But where can I find the statistical profile $\Prob_0: \BBB\to[0,1]$ characterizing the trustworthy behavior of the system $\SSS$? I could log the normal functioning of the system for a long time; but which observable system events $\BBB$ yield the relevant observations? 

The \emph{first limitation}\/ of scientific induction, discussed so far, is that it never proves, but only disproves its hypotheses. Here we confront its \emph{second limitation}: the null hypotheses cannot be extracted from the empiric data, but always have to be formulated \emph{a priori}.

\subsection{Compressing trust}\label{Sec:Apriori}
The problem of formulating \emph{a priori}\/ hypotheses was discussed in philosophy of science several centuries ago, but remained unsolved.  
The path towards the modern solutions was opened by Ray Solomonoff \cite{SolomonoffR:64}, and cleared by Andrei Kolmogorov \cite{KolmogorovA:statistic} and his school. The versions suitable for practical applications in machine learning and in statistics were developed by Jorma Rissanen \cite{RissanenJ:book}, Chris Wallace \cite{wallace2005statistical}, and many others. Very roughly, the idea is as follows.

Continuing with the notation from Sec.~\ref{Sec:Exp}, we still denote the set of hypotheses by $\Theta$. The problem is that we do not know the probabilities $\Prob_\theta :\BBB\to [0,1]$. We are, however, given a sufficiently large data sample, from which we extract the frequency distribution $\Prob : \BBB\to [0,1]$ of each observation.

The task is now to find a hypothesis $\theta = \theta_0\in \Theta$ such that $\Prob_0 : \BBB\to [0,1]$ maximizes the conditional probability $\Pr(\theta_0|x)$ in \eqref{Bayes} when the behavior $x\in \BBB$ is observed. Since
\bea
\Prob (x) & = & \sum_{\psi\in \Theta} \Prob_\psi(x)\Pr(\psi)
\eea
the Bayes' formula \eqref{Bayes} now becomes
\bea\label{Bayes-min}
\Pr(\theta | x) & = & \frac{\Prob_\theta(x)\Pr(\theta)}{\Prob(x)}
\eea
The null hypothesis $\theta_0$ gives the probability distribution $\Prob_0:\BBB \to [0,1]$ such that  for the observed $x$ holds $\Prob(\theta_0|x)\geq \Prob(\theta|x)$ for all $\theta \in \Theta$. Since the probability $\Pr(x)$ is given by the observed data, the task only depends on the unknown hypotheses $\theta \in \Theta$.

The idea used by Solomonoff, Kolmogorov and others is to apply Occam's razor here, and to postulate that \emph{the simplest hypotheses have the highest \emph{a priori} probability}. The idea is implemented by taking into account the \emph{lengths of the descriptions}\/ of the probabilities in \eqref{Bayes-min}. Using the optimal Shannon-Fano encodings \cite{Cover-Thomas}, we can write a number $p\in [0,1]$ using $-\log p$ bits. The task of maximizing \eqref{Bayes-min} now becomes the task of minimizing
\bear
-\log \Pr(\theta | x) & = & -\log \Prob_\theta(x) - \log \Pr(\theta) + \log \Prob(x)
\eear
Since $\Prob(x)$ is fixed, this means that
\bea\label{MDL}
\theta_0 & = & \underset{{\theta\in \Theta}}{{\rm argmin}}
\left\{ -\log \Prob_\theta(x) - \log \Pr(\theta)\right\}
\eea
This is equivalent to $\Prob_0(x) \cdot\Pr(\theta_0) \geq \Prob_\theta(x) \cdot\Pr(\theta)$ for all $\theta \in \Theta$, which picks $\theta_0$ to maximize the chance that $x$ is observed. This is what makes $\theta_0$ the best \emph{a priori}\/ null hypothesis. The minimality of the description length $-\log \Pr(\theta_0)$ means that $\theta_0$ is the simplest. The minimality of $-\log \Pr_0(x)$, or equivalently the maximality $\Pr_0(x)$, means that $x$ is the most likely prediction of $\theta_0$. The minimality of $-\log \Prob_0(x) - \log \Pr(\theta_0)$ means that $\theta_0$ is the simplest hypothesis among those that predict $x$. 

Instantiated to the realm of trust, \eqref{MDL} thus says that the best trust hypothesis is the one that provides the shortest description of my notion of trust, which fits the observations that I have made. 

The rapidly expanding research area of algorithmic learning and statistical inference is concerned not only with the effective computations of the \emph{a priori}\/ hypotheses, but also with the situations where the succinct descriptions of the data and the hypotheses need to be combined with empiric data. The right-hand side  of \eqref{MDL} is roughly Rissanen's Minimum Description Length (MDL) \cite{RissanenJ:book} of the distribution of the observed data $x$. Wallace's Minimum Message Length (MML)  \cite{RissanenJ:book} differs in the compression methods used. Kolmogorov's minimal sufficient statistic \cite{KolmogorovA:statistic} uses the optimal computable encodings as the compression method. The standard compression algorithms, e.g. based on the very efficient Lempel-Ziv algorithms \cite{Lempel-Ziv-1,Lempel-Ziv-2} are also often used, and give reasonable results. In any case, the best null hypothesis is the one which best compresses the observed data $x$, within some given family of compression algorithms. The underlying idea is that the better we understand the data, the better we compress them. 

Although these methods give somewhat degenerate results when applied to our toy examples from Sec.~\ref{Sec:Testing}, just slightly larger trust hypotheses show the intuitive meaning of \eqref{MDL} in the realm of trust. My trust hypothesis should be the simplest description of the desired behaviors which best approximates the observed behaviors of the tested system.

\section{Background and future work}\label{Sec:Final}

The main claim of this paper is that the methods of statistical inference, on which modern science has been built, can be used to analyze and secure trust. We close the paper relating this idea with the general context of trust research, and in particular with the existing application of statistical methods to trust testing in the framework of intrusion detection.   

The literature about trust is very extensive, as it is studied in psychology, social sciences, economics, game theory \cite{BergJ:reciprocity,buskens2002social,luhmann1979trust}. Even within the closely related security research communities, the word 'trust' is used in several different meanings \cite{BoydC:trust-survey}. The notion of trust used in this paper is based on \cite{PavlovicD:FAST10}. 

A quantitative analysis of the process of trust building was initiated in \cite{PavlovicD:FAST08}. The question of trust decisions was, however, avoided by reducing them to the preferences extracted from the trust ratings. The question of trust measurements was avoided by reducing them to user ratings and feedback, which are usually available in web commerce, but not in general. In system security,  the task of quantifying security in general and trust in particular becomes a problem \cite{BellovinS:brittle,WilliamsL:philosophies}. In the present paper, we did not consider the problem of quantifying trust \emph{a posteriori}, i.e. using the measurements of the past performance, but focused on the harder problem of formulating the trust hypotheses \emph{a priori}, i.e. before any empiric data are available. This problem arises even if the satisfactory methods for quantifying trust and security \emph{a posteriori}\/ are available, because the data are not always available. On the other hand, understanding how to express the \emph{a priori}\/ trust beliefs may also help in devising and validating the methods to quantify them \emph{a posteriori}.

The idea of \emph{statistical intrusion detection}, going back to Dorothy Denning \cite{DenningD:IDM} and her work with Peter Neumann at SRI in the 80s, can be viewed as an application of statistics to detect the subjects or the components that are not trustworthy. An early survey is \cite{LuntT:IDS}. The practices of intrusion detection have evolved a lot since those early days, and the rule based methods seem to have found broader applications than the statistical methods. One of the reasons often mentioned is the difficulty to control the false positives that arise when statistical tests are used to detect the intruders. We explained in Sec.~\ref{Sec:NullTrust} why the statistical methodologies suggest that trust testing should be based on taking a trustworthy behavior as the null hypothesis, and why testing for anomalies and the untrustworthy behaviors leads to the false positives that are harder to control, and to less reliable results overall. In statistics, proving that someone is not trustworthy is not equivalent to disproving that they are trustworthy. The general method for controlling the false positives when disproving trust is outlined in Sec.~\ref{Sec:Pow}. The false positives thus emerge as a hard problem in statistical intrusion detection because it tests for the intrusions, and not for trust. The reason is, of course, that the intruder profiles are much easier to come by than the trustworthy profiles. In the Sec.~\ref{Sec:Formulating}, we discussed the way to solve this problem using the methods of algorithmic learning. Whether that brief discussion explained or obscured the idea, there is very little doubt that at least a theoretical solution lies in this direction. But the practical work towards implementing such computation-based scientific methodologies on the concrete problems of trust lies ahead. 

\subsection*{Acknowledgements.} 
The comments and suggestions provided by Cormac Herley, Volodya Vovk and the anonymous referees helped me to improve the text. Some of their most interesting questions had to be left for future work.

\bibliography{ref-HotSoS,PavlovicD,games}
\bibliographystyle{plain}



\end{document}